\documentclass[12pt]{article}
\usepackage{graphicx}
\usepackage{eurosym}

\newcommand\pubnumber{NuPhys2017-Dracos}
\newcommand\pubdate{\today}

\def\iphc{IPHC, Universit\'e de Strasbourg, CNRS/IN2P3\\
F-67037 Strasbourg, France.}
\def\support{\footnote{On behalf of the ESS$\nu$SB Project.}}

\def\Title#1{\begin{center} {\Large #1 } \end{center}}
\def\Author#1{\begin{center}{ \sc #1} \end{center}}
\def\Address#1{\begin{center}{ \it #1} \end{center}}

\newcommand\pubblock{\rightline{\begin{tabular}{l} \pubnumber\\
         \pubdate  \end{tabular}}}
\newenvironment{Abstract}{\begin{quotation}  }{\end{quotation}}
\newenvironment{Presented}{\begin{quotation} \begin{center} 
             PRESENTED AT\end{center}\bigskip 
      \begin{center}\begin{large}}{\end{large}\end{center} \end{quotation}}
\def\Acknowledgements{\bigskip  \bigskip \begin{center} \begin{large}
             \bf ACKNOWLEDGEMENTS \end{large}\end{center}}

\def\beq{\begin{equation}}
\def\eeq#1{\label{#1}\end{equation}}
\def\eeqn{\end{equation}}

\def\beqa{\begin{eqnarray}}
\def\eeqa#1{\label{#1}\end{eqnarray}}
\def\eeqan{\end{eqnarray}}

\let\bar=\overbar

\def\Dslash{\not{\hbox{\kern-4pt $D$}}}
\def\dslash{\not{\hbox{\kern-2pt $\del$}}}

\def\msb{{\bar{\ssstyle M \kern -1pt S}}}

\begin{document}
\begin{titlepage}
\pubblock

\vfill
\Title{The European Spallation Source neutrino Super Beam}
\vfill
\Author{ Marcos Dracos\support}
\Address{\iphc}
\vfill
\begin{Abstract}
After measuring in 2012 a relatively large value of the neutrino mixing angle $\theta_{13}$, the door is now open to observe for the first time a possible CP violation in the leptonic sector.
The measured value of $\theta_{13}$ also privileges the 2nd oscillation maximum for the discovery of CP violation instead of the usually used 1st maximum.
The sensitivity at this 2nd oscillation maximum is about three times higher, with a lower influence of systematic errors, than for the 1st maximum.
Going to the 2nd oscillation maximum necessitates a very intense neutrino beam with the appropriate energy.
The world's most intense pulsed spallation neutron source, the European Spallation Source, will have a proton linac with 5 MW power and 2 GeV energy.
This linac, under construction, also has the potential to become the proton driver of the world's most intense neutrino beam with high probability to discover a neutrino CP violation.
The physics performance of that neutrino Super Beam in conjunction with a megaton underground Water Cherenkov neutrino detector installed at a distance of about 500~km from ESS has been evaluated.
In addition, the choice of such detector will extent the physics program to proton-decay and astrophysics searches.
The ESS proton linac upgrades, the accumulator ring needed for proton pulse compression, the target station, the far detector and the physics potential are described.
In addition to neutrinos, this facility will also produce at the same time a copious number of muons which could be used by a low energy neutrino facility for sterile neutrino searches, a future Neutrino Factory or a Muon Collider.
The ESS neutron facility will be fully ready by 2023 at which moment the upgrades for the neutrino facility could start.

\end{Abstract}
\vfill
\begin{Presented}
NuPhys2017, Prospects in Neutrino Physics\\
Barbican Centre, London, UK,  December 20--22, 2017
\end{Presented}
\vfill
\end{titlepage}
\def\thefootnote{\fnsymbol{footnote}}
\setcounter{footnote}{0}

\section{Introduction}

The CP violation observed in the hadronic sector is by several orders of magnitude too low to explain the matter-antimatter asymmetry observed in the Univers.
Other sources of CP violation are needed to explain this phenomenon as a CP violation in the leptonic sector.
The measurement of CP violating parameter $\delta_{CP}$ appearing in the neutrino mixing matrix, taking into account the relatively high value of the last measured mixing angle $\theta_{13}$, is now accessible to long baseline neutrino experiments.

For this observation, very high intensity neutrino beams are needed.
Several proposals have been placed proposing to work as usual on the 1st oscillation maximum, mainly for statistical reasons.
The success of these projects strongly depends on the systematic errors, which values will definitely be known at the end of the project.
To mitigate this problem, one possibility is to work on the 2nd oscillation maximum where the dependence on systematic errors is significantly lower compared to the 1st oscillation maximum~\cite{Nunokawa:2007qh, Coloma:2011pg, Parke:2013pna}.
Moreover, the sensitivity to matter- antimatter asymmetry is higher at the 2nd oscillation maximum since its value is proportional to $0.75\sin\delta_{CP}$, instead of $0.3\sin\delta_{CP}$ at the 1st oscillation maximum.
The drawback is that significantly higher statistics are needed.

For high statistics a very intense neutrino beam is needed.
This is what is proposed by the European Spallation Source (ESS) neutrino Super Beam (ESS$\nu$SB)~\cite{Baussan:2013zcy}, by using the ESS 5~MW/2~GeV proton linac and observing the oscillation $\nu_\mu \rightarrow\nu_e$.
For the proposed neutrino facility, a far detector could be placed in an existing mine in the north of Lund at a distance of about 500~km.

The ESS neutron facility is under construction in Lund, Sweden since 2014 and it will be ready by 2023.

\section{The ESS facility}

\begin{figure}[htb]
\centering
\includegraphics[width=0.9\textwidth]{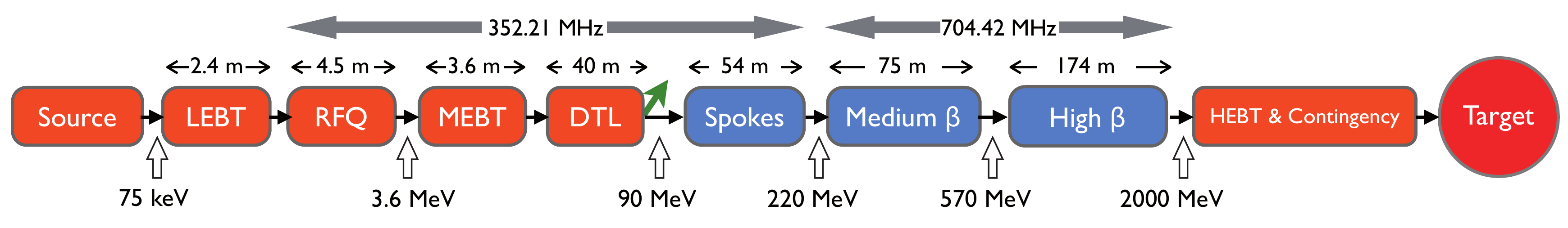}
\caption{\small Schematic view of the ESS proton linac.}
\label{linac}
\end{figure}

The ESS facility is devoted to neutron activities related to material irradiations.
For this, a very powerful proton linac is used with the protons hitting a target to produce neutrons.
The characteristics of this proton linac are summarised by Table~\ref{linac}.
The proton kinetic energy will be 2.0~GeV with the possibility to go up to 3.6~GeV using empty space at the end of the linac (Fig.~\ref{figlinac}).

\begin{table}[htp]
\begin{center}
\begin{tabular}{l|c}
Parameter & Value \\
\hline
Average beam power & 5~MW \\
Peak beam power & 125~MW \\
Proton kinetic energy & 2.0 GeV \\
Average macro--pulse current & 62.5~mA \\
Macro--pulse length & 2.86~ms \\
Pulse repetition rate & 14~Hz \\
Annual operating period & 5000~h \\
\hline
\end{tabular}
\caption{\small Main ESS facility parameters of the proton beam.}
\label{figlinac}
\end{center}
\end{table}

The linac will be pulsed at a frequency of 14~Hz with a pulse length of 2.86~ms.
The duty cycle is only 4\%, the rest of the time the linac being empty.
By doubling the pulse rate and keeping the same instantaneous beam power, another 5~MW averaged power can be obtained to be used by a neutrino facility.
Under these assumptions $2.7\times 10^{23}$ protons on the target can be obtained per year.


\section{Neutrino Beam}

\begin{figure}[hbt]
\centering
\includegraphics[width=0.8\textwidth]{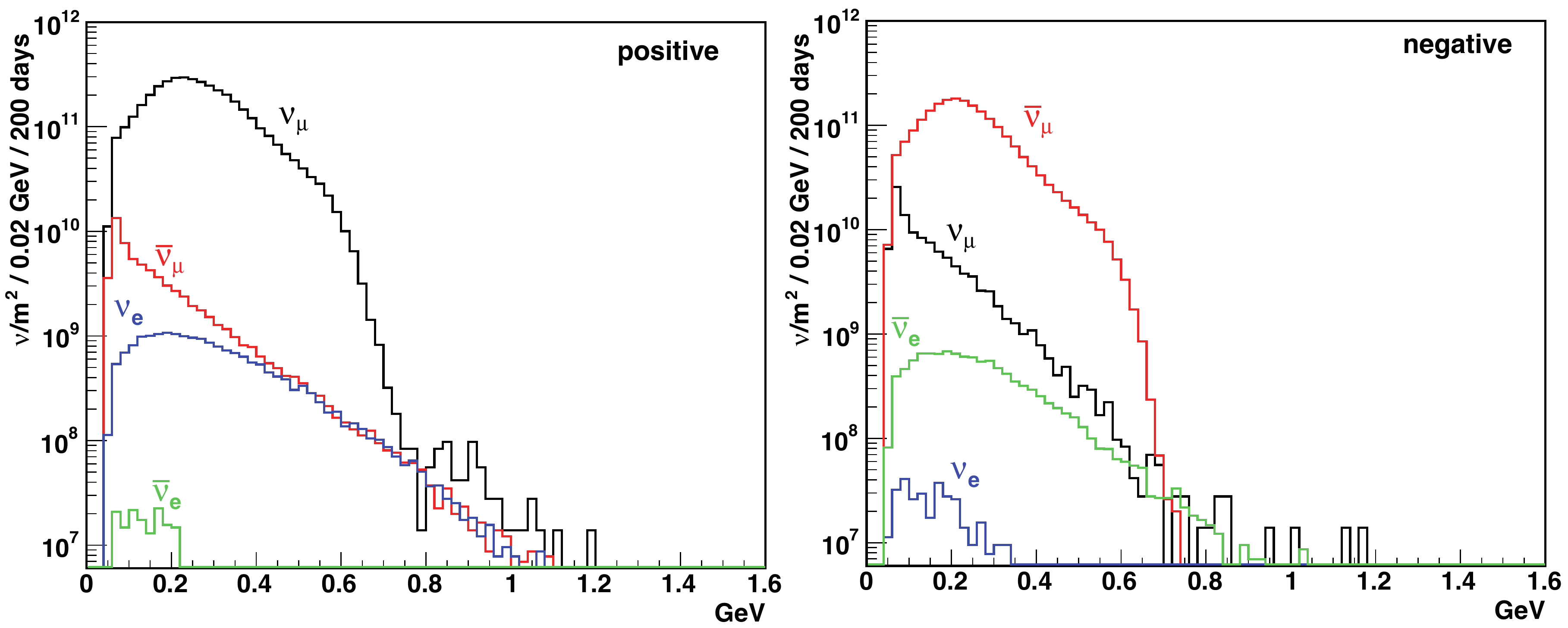}
\caption{\small Neutrino energy distribution at a distance of 100~km on--axis from the target station, for 2.0~GeV protons and positive (left, neutrinos) and negative (right, anti--neutrinos) horn current polarities, respectively.}
\label{fluence}
\end{figure}

Using the ESS linac proton beam, a neutrino beam can be produced using an appropriate target and a hadron collecting system.
By using a target station similar to the one proposed by the EU FP7 EURO$\nu$ Design Study~\cite{1305.4067, 1212.0732} and 25~m length hadron decay tunnel, the neutrino/antineutrino spectrum of Fig.~\ref{fluence} can be obtained per year (200 days).
Preliminarily, it is supposed that the facility will be operated during two years in neutrino mode (positive current in the magnetic horn) and eight years in antineutrino mode (negative current) in order to detect in the far detector the same number of neutrinos than antineutrinos.

\begin{table}[htp]
\begin{center}
\begin{tabular}{|c|c|c|c|c|}\hline
 & \multicolumn{2}{c|}{positive} & \multicolumn{2}{c|}{negative} \\
\cline{2-5} & $N_\nu\ (\times 10^{10})$/m$^2$ & \% &$N_\nu\ (\times 10^{10})$/m$^2$ & \% \\
\hline $\nu_\mu$           & 396 & 97.9     & 11   & 1.6 \\
\hline $\bar\nu_\mu$ & 6.6   & 1.6     & 206 & 94.5 \\
\hline $\nu_e$               & 1.9   & 0.5     & 0.04   & 0.01 \\
\hline $\bar\nu_e$     & 0.02   & 0.005 & 1.1   & 0.5 \\
\hline
\end{tabular}
\caption{\small Number of neutrinos per~m$^2$ crossing a surface placed on--axis at a distance of 100~km from the target station  during 200 days for 2.0~GeV protons and positive/negative horn current polarities.}
\label{tabfluence}
\end{center}
\end{table}

From Table~\ref{tabfluence} it can be seen that a relatively pure muon neutrino beam can be obtained.
Since for the CP violation observation the oscillation $\nu_\mu \rightarrow\nu_e$ will be used, any presence of electron neutrinos in the primary neutrino beam will disturb this study.
The primary electron neutrino contribution is only 0.5\%, low enough to significantly affect this study.
This contribution could be used by a near detector to study the electron neutrino cross-section at the relevant energies of this project.
For this purpose, a strong rejection of muon neutrino and neutral current events has to be obtained.

\section{Far detector}

For the ESS$\nu$SB studies the Water Cherenkov MEMPHYS detector~\cite{deBellefon:2006vq} already proposed by EURO$\nu$ and LAGUNA-LBNO~\cite{Agarwalla:2014tca} EU projects has been adopted.
This far detector could be placed in or near an active mine with suitable rock characteristics.
Two mines have been identified in Sweden suited to host this detector.
The first mine, Zinkgruvan, is located at a distance of 360~km from Lund, while the second one, Garpenberg, at 540~km.

MEMPHYS, with its 500~kt fiducial volume (two cylinders of 65~m diameter and 100~m height), could also be used for proton decay searches, super nova neutrino detection, solar and atmospheric neutrinos.
In these studies, the performance of MEMPHYS, as reported in~\cite{1206.6665}, has been used.
The $\nu_e$ detection is of the order of 50\% for a PMT photocathode coverage of 30\%.
This coverage could now be increased by keeping the same price by using MCP-PMTs, as those used by the JUNO experiment~\cite{An:2015jdp}.
This option is now under investigation.

\begin{figure}[hbt]
\centering
\includegraphics[width=0.9\textwidth]{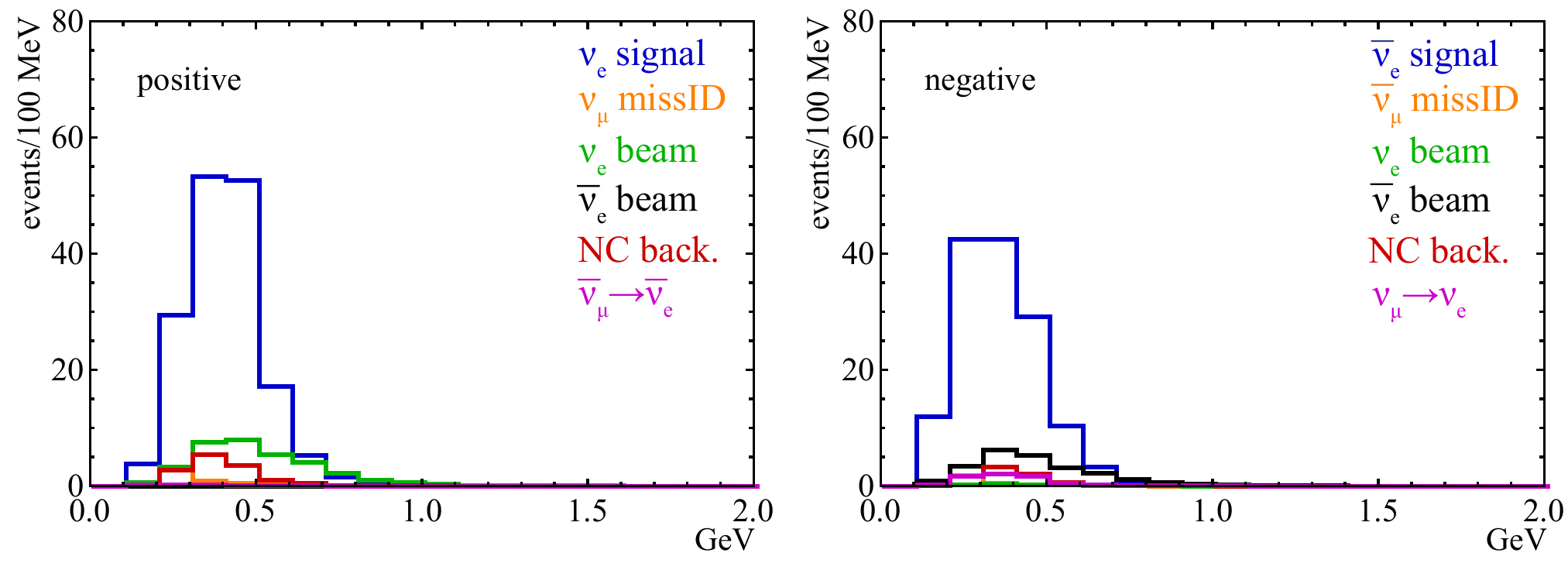}
\caption{\small Energy distributions of the detected electron neutrinos (positive) and anti--neutrinos (negative) including background contribution as reconstructed by MEMPHYS detector for two years of neutrino running (left) plus eight years of antineutrino running (right) and a baseline of 540~km (2.0~GeV protons, $\delta_{CP}=0$).}
\label{detected}
\end{figure}

Fig.~\ref{detected} shows the detected neutrino energy distribution for a far detector placed in Garpenberg mine (540~km).
The first (second) plot presents the distribution for neutrino two (eight) years run.
For both modes, neutrino/antineutrino, about 300 $\nu_e$'s coming from the muon neutrino oscillation are expected to be detected.
The mean neutrino energy is of the order of 0.4~MeV, region around which the neutrino interaction cross-sections drops down very quickly.
On the other side, almost only quasi-elastic events are expected avoiding $\pi^0$ production in the resonant regime.
The absence of $\pi^0$'s explain the low background of Fig.~\ref{detected}.

\begin{figure}[hbt]
\centering
\begin{minipage}{0.43\linewidth}
\includegraphics[width=0.99\textwidth]{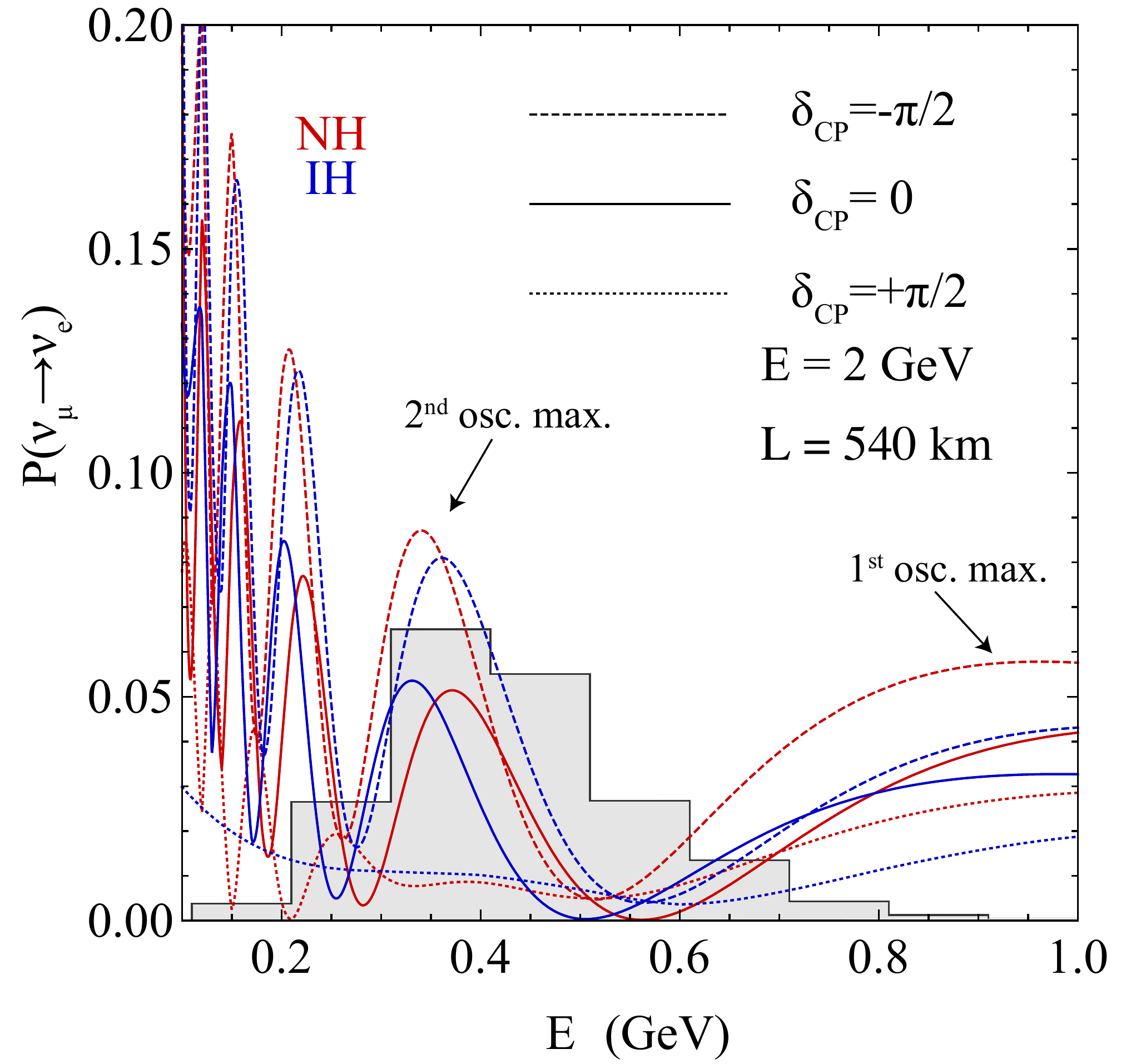}
\caption{\small $\nu_\mu \rightarrow \nu_e$ oscillation probability as a function of the energy. The solid (dashed) lines are for normal hierarchy (inverted). The shaded histogram is the energy distribution of electron neutrinos detected by the far detector.}
\label{secosc}
\end{minipage}\hspace{2pc}%
\begin{minipage}{0.51\linewidth}
\includegraphics[width=0.99\textwidth]{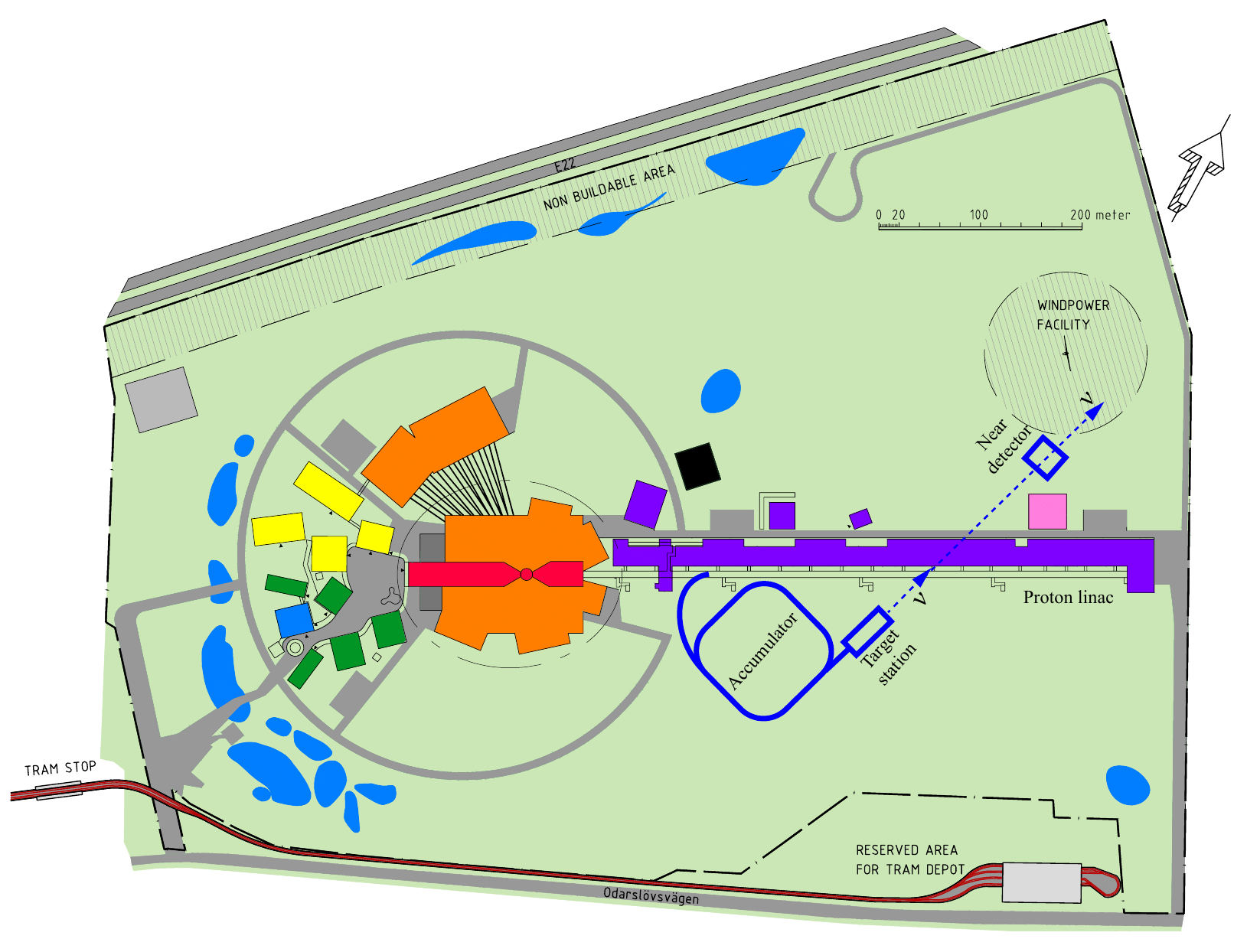}
\caption{\small Layout of the ESS installation with a possible neutrino facility implementation (accumulator, target station, near detector).}
\label{layout}
\end{minipage} 
\end{figure}

As said above, the aim of the ESS$\nu$SB project compared to other long baseline projects, is to use the second oscillation maximum to observe a CP violation in the neutrino sector.
Fig.~\ref{secosc} presents the $\nu_\mu \rightarrow \nu_e$ oscillation probability as a function of the neutrino energy for normal and inverted mass hierarchies, for a far detector placed in the Garpenberg mine.
The shaded distribution is the $\nu_e$ distribution of Fig.~\ref{detected}, just to show that the second oscillation maximum of the oscillation probability is well covered.

\section{ESS upgrade}

Few upgrades are needed to the ESS facility to add on top of the neutron facility the neutrino one.
The ESS proton linac pulses of 2.86~ms are too long for the neutrino facility.
This because of the high current to be sent to the magnetic horn and considerable Joule effect produced in this hadron collecting device placed in the target station.
For this reason and also to decrease the cosmic ray background in the far detector, shorter pulses, of the order of few $\mu$s, are needed.
This could also be benefit to the neutron facility because this modification could significantly increase the neutron brightness.

To shorten the proton pulses, as for SNS in the USA, an accumulation and compressing ring is needed.
In order not to go out of the already allocated area of ESS, a ring with a circumference of bout 400~m is proposed, reducing the proton pulses to about 1.5~$\mu$s.
Due to space-charge effects at the entrance of the accumulator, H$^-$ ions have to be used, which would be stripped just before entering the ring.
This implies that H$^-$, instead of protons, have to be produced and accelerated in the ESS linac.
The modifications needed to accelerate H$^-$ have already been studied.
While no showstoppers have been identified, the cost of these modifications is not negligible and is of the order of 250~M\euro (including a 2.5~GeV proton energy upgrade).
All these studies can be found in a CERN report~\cite{cern}.

A target station is needed at the exit of the accumulator to produce the neutrino beam.
For this, ESS$\nu$SB has adopted the design already proposed by EURO$\nu$.
In this design, taking into account the relatively low proton energy, the target is placed inside the magnetic horn.
In order to mitigate the effect of the very intense proton beam, a four target/horn system is proposed.
To keep the system as simple as possible, no reflector is used.
Under these conditions, each target is supposed to receive 1/4 of the total 5~MW proton power.
For cooling reasons, each target is composed of a canister fulfilled with titanium alloy spheres (few mm diameter).
In this way, He gas can circulate around the spheres to remove the heat and potential vibration problems could also be avoided.

The decay tunnel, after the target/horn system, has a length of the order of 25~m and could also be filled with He gas in order to avoid intermediate windows.
A graphite beam dump would stop all the remaining protons, pions and muons.

A near detector can be placed after the target station to monitor the neutrino beam but also to measure neutrino cross-sections.
The layout of the whole ESS facility with a possible neutrino facility implementation is depicted by Fig.~\ref{layout}.

\section{Physics performance}

\begin{figure}[h]
\centering
\begin{minipage}{0.40\linewidth}
\includegraphics[width=0.99\textwidth]{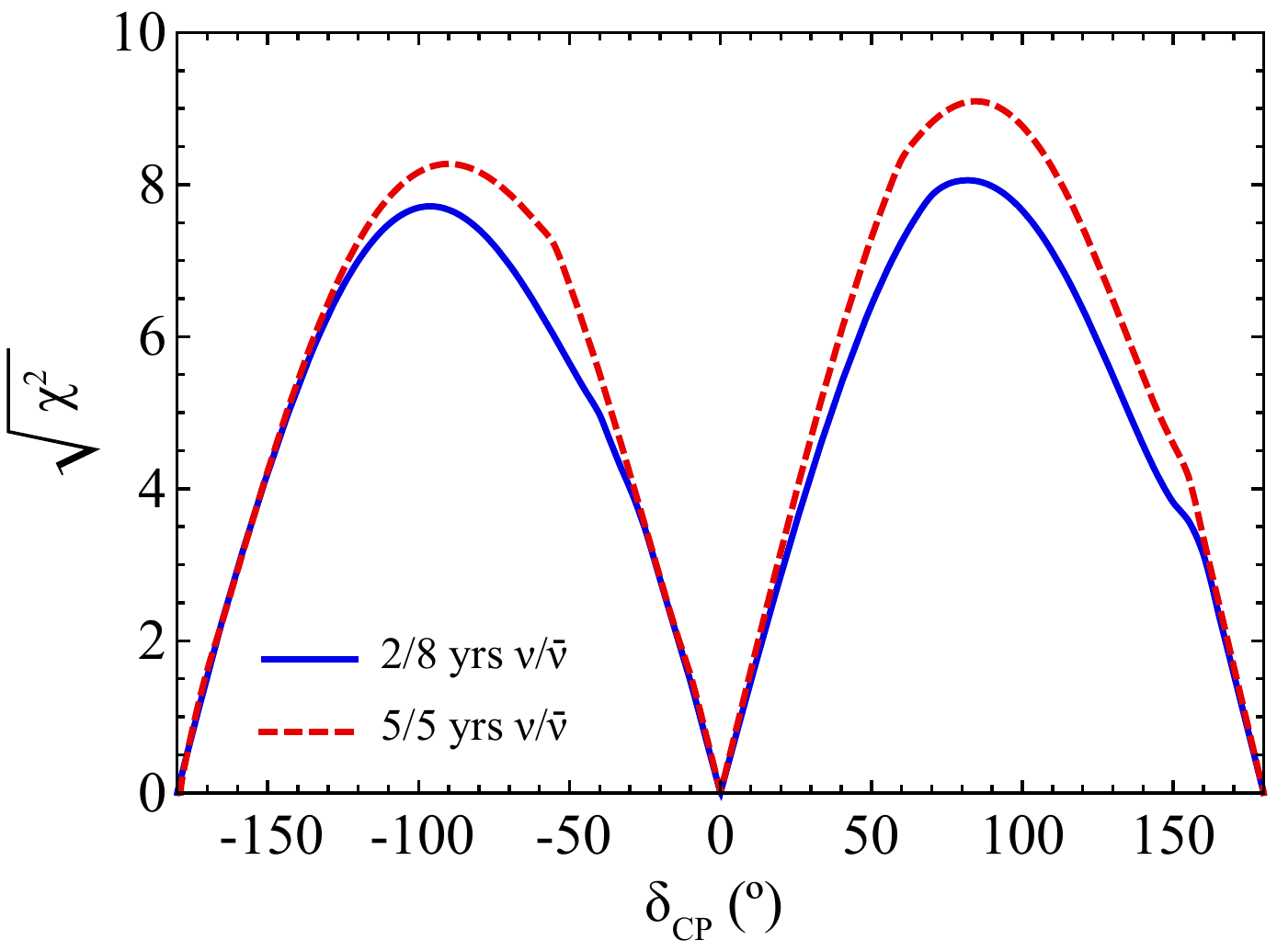}
\caption{\label{cpv}\small The significance to discover a CP violation as a function of the $\delta_{CP}$ parameter.}
\end{minipage}\hspace{2pc}%
\begin{minipage}{0.40\linewidth}
\includegraphics[width=0.99\textwidth]{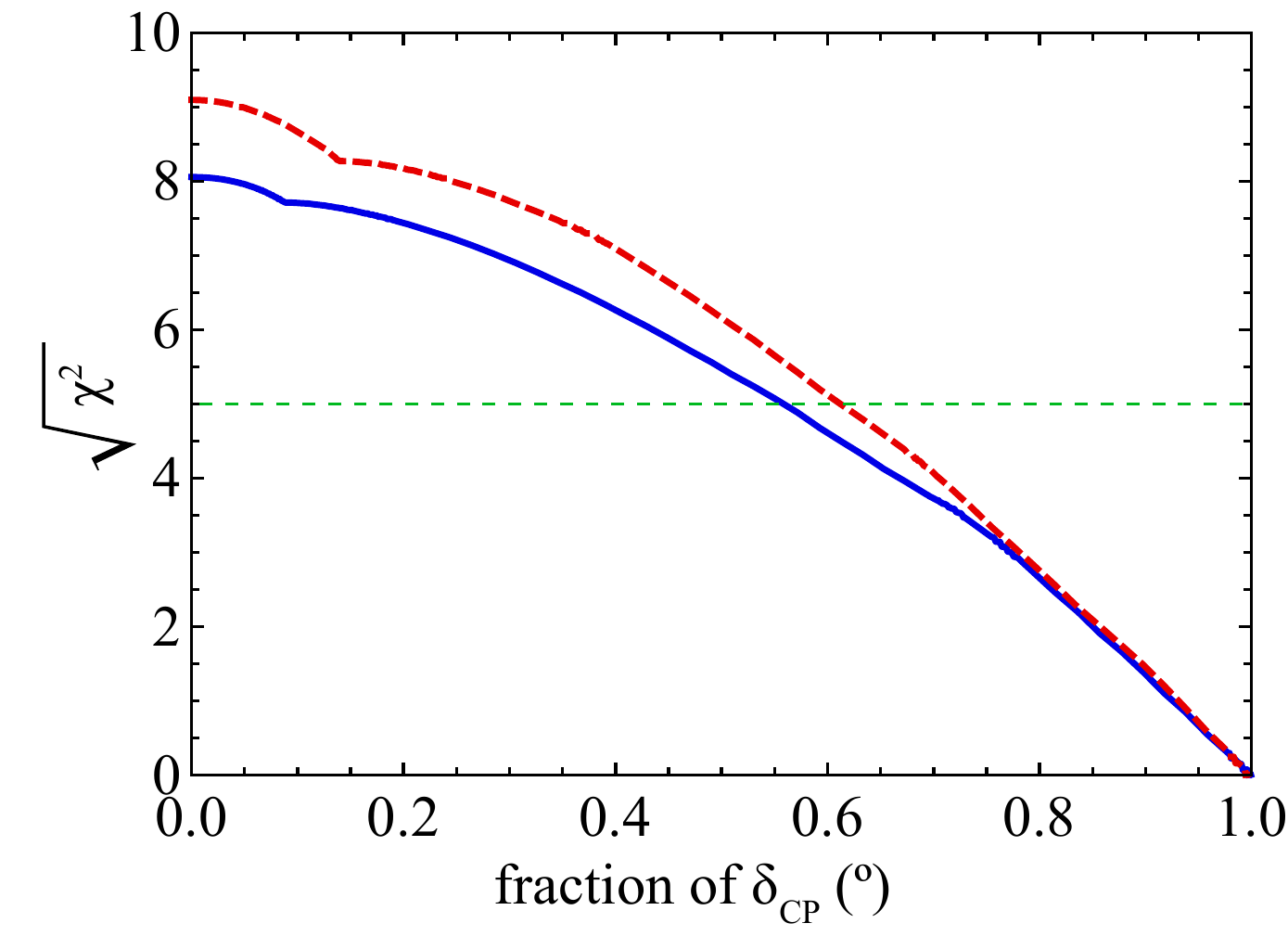}
\caption{\label{cpf}\small The significance with which CP violation can be discovered as function of the fraction of the full~$\delta_{CP}$ range.}
\end{minipage} 
\end{figure}

For the evaluation of the physics performance of the proposed facility, the main parameters defined by EURO$\nu$ have been used.
The baseline considered is the one corresponding to the Garpenberg mine (540~km).
Fig.~\ref{cpv}~\cite{enrique} presents the significance versus $\delta_{CP}$ to discover CP violation, while Fig.~\ref{cpf} shows the significance as a function of the covered $\delta_{CP}$ range.
For a 5~$\sigma$ discovery, a coverage of about 60\% of  $\delta_{CP}$ values is obtained.
For these evaluations the systematic errors reported in~\cite{1209.5973} have been used (mainly 5\%/10\% systematics for signal/background).

Further optimisations engaged now in the approved EU H2020 Design Study ESS$\nu$SB could improve this performance on CP violation discovery.
This EU project has started the 1st of January 2018 and will last for four years.
At the end of this study a Conceptual Design Report (CDR) of the whole neutrino facility will be delivered.


On top of the neutrino production, this facility will also produce a huge number of muons.
At the level of the beam dump, it is expected to collect about $16\times 10^{20}$ muons per year with a mean momentum of 0.5~GeV.
These muons could be used for muon cooling R\&D, sterile neutrino searches, a future Neutrino Factory or a muon collider. 

\section{Present context}

The ESS$\nu$SB project is now supported by the COST Action EuroNuNet and a EU H2020 Design Study, which will end by 2021.
By this time ESS$\nu$SB will deliver a CDR of the overall proposed neutrino installation.
After this period it is planed to have a preparatory period up to 2024 to produce a Technical Design Report and perform R\&D on subjects defined by the previous studies.
By this time the ESS neutron facility will be ready for running.

A preparatory phase between 2025 and 2026 can follow this period before starting implementation of the neutrino facility and construction of the far detector.
The whole construction of the neutrino facility is expected to last from 2027 to 2033 with one or two years commissioning after.
Under these assumptions, the data taking period could start around 2036.

\section{Conclusions}

The sensitivity to the CP violating parameter $\delta_{CP}$ is significantly higher on the 2nd maximum of the $\nu_\mu \rightarrow\nu_e$ oscillation probability.
Moreover,  the neutrino long baseline projects operated at the 2nd oscillation maximum suffer less by systematic errors.

It is demonstrated that the use of the ESS linac can produce a very intense neutrino beam allowing the operation of this facility at the 2nd oscillation maximum.
Under these conditions, more than 50\% of the $\delta_{CP}$ range can be covered with $5\ \sigma$ confidence level to discover a CP violation in the leptonic sector.

Preliminary feasibility studies have not identified showstoppers.
Upgrades of the ESS neutron facility have been proposed in order to add the neutrino facility.
This includes the addition of a compressor ring and of a new target station.

The proposed far detector has also a rich astroparticle physics programme which could start before the neutrino facility is accomplished.
On top of this, the large production of muons together with neutrinos, could allow in a later stage to transform this installation to a muon facility for a Neutrino Factory or/and a muon collider.

The proposed second generation long baseline project is expected to start taking data by 2036.

\Acknowledgements

This project is supported by the COST Action CA15139 ``Combining forces for a novel European facility for neutrino-antineutrino symmetry-violation discovery'' (EuroNuNet).
It has also received funding from the European Union's Horizon 2020 research and innovation programme under grant agreement No 777419.

\end{document}